\documentclass[numberedappendix]{emulateapj}
\usepackage{amstext}
\usepackage{apjfonts}
\usepackage{amsmath}
\usepackage{amssymb}
\usepackage[colorlinks=true,linkcolor=blue,citecolor=blue]{hyperref}
\usepackage{comment}

\begin{document}

\title{Extracting Periodic Transit Signals from Noisy Light Curves using Fourier Series} 

\author{Johan Samsing$^{1,2}$} 
\altaffiltext{1}{Department of Astrophysical Sciences, Princeton University, Peyton Hall, 4 Ivy Lane, Princeton, NJ 08544, USA}
\altaffiltext{2}{Einstein Fellow}

\begin{abstract} 

We present a simple and powerful method for extracting transit signals associated with a known transiting planet from noisy light curves.
Assuming the orbital period of the planet is known and the signal is periodic, we illustrate that systematic noise can be removed in Fourier space at all frequencies,
by only using data within a fixed time frame with a width equal to an integer number of orbital periods.
This results in a reconstruction of the full transit signal which on average is unbiased, despite that no prior knowledge of either the
noise or the transit signal itself is used in the analysis. The method has therefore clear advantages over standard phase folding, which normally requires
external input such as nearby stars or noise models for removing
systematic components. In addition, we can extract the full orbital transit signal ($360$ degrees) simultaneously,
and \emph{Kepler} like data can be analyzed in just a few seconds.
We illustrate the performance of our method by applying it to a dataset composed of light
curves from \emph{Kepler} with a fake injected signal emulating a planet with rings.
For extracting periodic transit signals, our presented method is in general the optimal and least biased estimator
and could therefore lead the way toward the first detections of, e.g., planet rings and exo-trojan asteroids.

\end{abstract}


\section{Introduction}

Our understanding of where exoplanets form, their orbital distribution, and even properties such as mass and radius
has greatly improved over recent years \citep{2014Natur.513..336L, 2015arXiv150105685L}.
Several methods have contributed to this progress, from mass sensitive measurements including pulsar timings \citep{1992Natur.355..145W}, radial velocimetry
\citep{1995Natur.378..355M, 2014Natur.513..328M} and microlensing \citep{2004ApJ...606L.155B}, to transit
photometry \citep{2000ApJ...529L..41H, 2000ApJ...529L..45C}, which in turn depends on the planet radius.
Recent developments have primarily been driven by photometric surveys
operating from both the ground (OGLE \citep{2002AcA....52....1U}, HAT \citep{2002PASP..114..974B} and WASP \citep{2006PASP..118.1407P})
and space (CoRoT \citep{2009A&A...506..411A}, \emph{Kepler} \citep{2010Sci...327..977B}).
The emerging picture is that exoplanets are not rare, but exist with a wide variety of masses, orbital sizes, and eccentricities,
and often in multiplanet configurations which can be very different from our own solar system \citep{2015arXiv150105685L}.
Despite the success in detecting planets, there are still major gaps in our understanding
of how they form and their subsequent evolution. The problem covers all aspects of astrophysics
from the formation and evolution of the protoplanetary disk (see e.g., \cite{2014prpl.conf..411T}), to late time
gravitational dynamics including both strong field planet-planet scattering and secular evolution \citep{2012RAA....12.1044B}.
Due to this complexity, observations play a crucial role in driving the field.

In this paper we focus on photometric measurements
and how to improve the scientific output from transit light curves.
One of the present challenges is to reduce the noise to
a level where 'higher-order photometric effects' \citep{2011exha.book.....P}, with 
relative flux variations of order $10^{-4}$, can be resolved. This level of precision could lead to the first detection of
exomoons \citep{2012MNRAS.419..164S, 2012ApJ...750..115K, 2014ApJ...787...14H, 2015arXiv150205033H},
planet rings \citep{2004ApJ...616.1193B}, and exo-trojan asteroids \citep{2002AJ....124..592L}, as well as teach us about the
rotation of the transiting planets \citep{2014ApJ...796...67Z} and even their atmosphere \citep{2002ApJ...572..540H, 2010ApJ...720..904S}, to name a few examples.
However, such variations are very hard to resolve with present observations and noise removal techniques.

Motivated by these challenges, we
here present a fast and effective method for extracting small flux variations associated with a known planet transit, from light
curves with random and systematic noise. The noise represents here all components which are not associated with the transit signal itself such
as instrumental noise and real stellar flux variations.
By assuming the transit signal is periodic, we illustrate how one can
remove systematic noise in a model independent way, which makes it possible to
resolve finer flux variations than standard phase folding techniques\footnote{Phase folding techniques are in this paper loosely referring
to methods which use out-of-transit fits and stacking in real space to reduce systematic and random noise, respectively.}. 
This could directly improve the ongoing search for planetary rings and trojan-asteroids, as well as 
lead to better measurements of any phenomena associated with the planet transit itself.

The new strategy we propose is based on removing systematic noise in Fourier space. We show this becomes
especially effective if one only uses data within a fixed time frame starting from the middle of the first planet transit and ending in the last.
In this way any signal associated with the transiting planet will be constrained to discrete and separate frequencies in Fourier space, which
makes the noise separation possible.
Our method does not rely on any prior knowledge of either
the noise or signal, which also makes it fast enough that several quarters of \emph{Kepler} data
can be analyzed in seconds.
If the noise is not directly correlated with the periodic signal,
our method will, on average, return an unbiased reconstruction of the full orbital transit signal.
Fourier transforms have previously been used for, e.g., finding short period planets \citep{2014ApJ...787...47S}, as well as
removing noise in exoplanet emission spectra \citep{2010Natur.463..637S, 2014ApJ...796...48Z}, but has to our knowledge not been applied for
extracting the transit shape itself in the way we propose.

This paper is organized as follows. In Section \ref{sec:Transit_Signals_and_Light_curve_Observations} 
we describe how the observed flux, possible signal, and noise are related. The notation and assumptions from this
section will be used in the remaining parts. In Section \ref{sec:Reconstruction_of_Periodic_Transit_Signals} we describe our new Fourier
method step by step and how it can be applied to reconstruct any periodic transit shape. The performance of the method is shown in
Section \ref{sec:Numerical_Example}, where we apply it to reconstruct a fake signal injected into real \emph{Kepler} light curves.
Conclusions are given in Section \ref{sec:Conclusions}.


\section{Transit Signals and Light curve Observations}\label{sec:Transit_Signals_and_Light_curve_Observations}

We consider a planetary system consisting of a star with a single planet.
If observed from the right position, the planet will
move in front of the star during its orbit, which gives rise to periodic dips in the observed light curve.
We denote this the main transit.
Objects associated with the planet such as rings, moons, and trojan-asteroids
can likewise block parts of the stellar light, and can therefore be indirectly seen as
small flux variations on top of the main transit signal.
A few examples are shown in Figure \ref{fig:Ill_planet_transit}.
In this section we describe the assumptions we make about how the observed flux, transit signal, and noise are related to each other.

\subsection{Flux from a Star with an Orbiting Planet}\label{sec:Flux_from_a_Star}

For this paper we consider observations of a star with one transiting planet.
The observed flux $F_{\text{obs}}$ can in the majority of cases be modeled as
\begin{equation}
F_{\text{obs}} = \bar{F_{*}} + \Delta{F} + \bar{F_{*}}I^{\mathcal{S}},
\label{eq:Fobs_def}
\end{equation}
where $\bar{F_{*}}$ is the mean flux from the star, $\Delta{F}$ is considered noise, and $I^{\mathcal{S}}$ is the fractional flux modulation coursed by
orbiting objects including the planet. The modulation term $I^{\mathcal{S}}$ is typically
negative, but reflections from the planet surface can give rise to
positive modulations. The noise term $\Delta{F}$ is assumed to be small compared to $\bar{F_{*}}$,
and represents all systematic and random flux variations which are not associated with the true transit signal.
By defining the observable quantity
$I(t) \equiv F_{\text{obs}}/\bar{F_{*}} - 1$ and the relative noise contribution $I(t)^{\mathcal{N}} \equiv \Delta{F}/\bar{F_{*}}$, we can rewrite
equation \eqref{eq:Fobs_def} as
\begin{equation}
I(t) = I(t)^{\mathcal{S}} + I(t)^{\mathcal{N}}.
\label{eq:I_sum_IN_IS}
\end{equation}
In the remaining part of the paper we generally refer to $I(t)$ as the data, $I(t)^{\mathcal{S}}$ as the signal and $I(t)^{\mathcal{N}}$ as the noise.
In the following sections we show that if the signal $I(t)^{\mathcal{S}}$ is periodic, 
then one can make a precise reconstruction of its transit shape by working in Fourier space. 


\section{Reconstruction of Periodic Transit Signals}\label{sec:Reconstruction_of_Periodic_Transit_Signals}

We now present our new method for extracting periodic transit signals. The method relies on the basic assumption that the transit
signal must be strictly periodic, with the same period as the orbiting planet.
Besides that, there are no restrictions on the shape of the signal, i.e., the method can be used to reconstruct
transits with any shape.
The assumed periodicity makes it possible to work in Fourier space, which enables us to
remove systematic noise components. This is in contrast to standard phase folding techniques where the real space
stacking is only optimal for reducing random noise.
In the sections below we describe our proposed Fourier method step by step. 

\subsection{Representing Data as a Fourier Series}

The general strategy is to extract the periodic transit signal in Fourier space.
We therefore first consider the data $I(t)$ represented as a Fourier series
\begin{equation}
I(t) = \sum_{n=1}^{N/2} a_{n}{\text{cos} \left( \frac{2{\pi}n(t-\tau_{i})}{|\tau_{f} - \tau_{i}|} \right)} + \sum_{n=1}^{N/2} b_{n}{\text{sin} \left( \frac{2{\pi}n(t-\tau_{i})}{|\tau_{f} - \tau_{i}|} \right)},
\label{eq:I_Fourier_ex}
\end{equation}
where $t$ is time, $N$ is the number of data points, and $\tau_{i}$, $\tau_{f}$ are denoting the starting time
and ending time of the dataset used in the analysis, respectively. We refer to this fixed time frame as the 'data window'
and denote its width, $|\tau_{f}-\tau_{i}|$, by $T_{w}$. The importance of the width and location of this window is
discussed in section \ref{sec:Choosing_the_Optimal_Data_Window}.
From our signal and noise model shown in equation \eqref{eq:I_sum_IN_IS} and by the linearity of the Fourier transformation, the Fourier
coefficients can be divided up into signal and noise as
\begin{equation}
a_{n} = a_{n}^{\mathcal{S}} + a_{n}^{\mathcal{N}}, \ b_{n} = b_{n}^{\mathcal{S}} + b_{n}^{\mathcal{N}}.
\label{eq:Fourier_coefficients_signal_plus_noise}
\end{equation} 
The idea is now to separate the signal and the noise in the Fourier domain, as
shown in equation \eqref{eq:Fourier_coefficients_signal_plus_noise}.
In the section below we show that a specific choice for $\tau_{i}$ and $\tau_{f}$ will make this approach possible.

\begin{figure}
\includegraphics[width=\columnwidth]{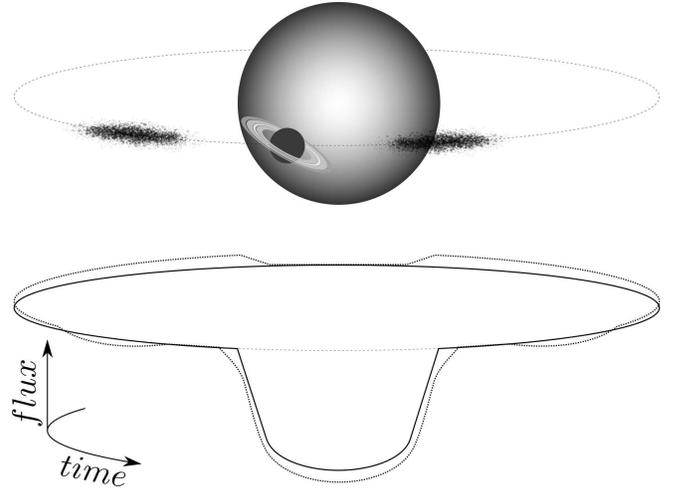}
\caption{Illustration of a planet with rings and a population of exo-trojan asteroids.
The \emph{top} plot shows a schematic drawing of the system, where the \emph{bottom} plot shows the corresponding
transit signal. The illustration is not to scale, but intends to illustrate how a typical signal could look like and especially how
planet rings and asteroids contribute to the signal.
A real signal has encoded much more information such as how fast the planet rotates and if it has an atmosphere,
to name a few examples. However, it is in general very hard to resolve the resultant flux variations
which would show up as small perturbations to the main transit (solid line) at the percent level.
In this paper we present a way of improving this by working in Fourier space.}
\label{fig:Ill_planet_transit}
\end{figure}

\subsection{Choosing the Optimal Data Window}\label{sec:Choosing_the_Optimal_Data_Window}

The Fourier coefficients obtained from using the full data set $I(t)$
will in general have broad features, and the underlying signal will overlap with the
noise spectrum, even if the signal has strictly periodic components.
This is mainly due to 'window effects' where the position and width of the data window interfere with the
Fourier modes of the data. Separating the coefficients as illustrated in equation \eqref{eq:Fourier_coefficients_signal_plus_noise}, is
therefore not possible in general.

\begin{figure}
\includegraphics[width=\columnwidth]{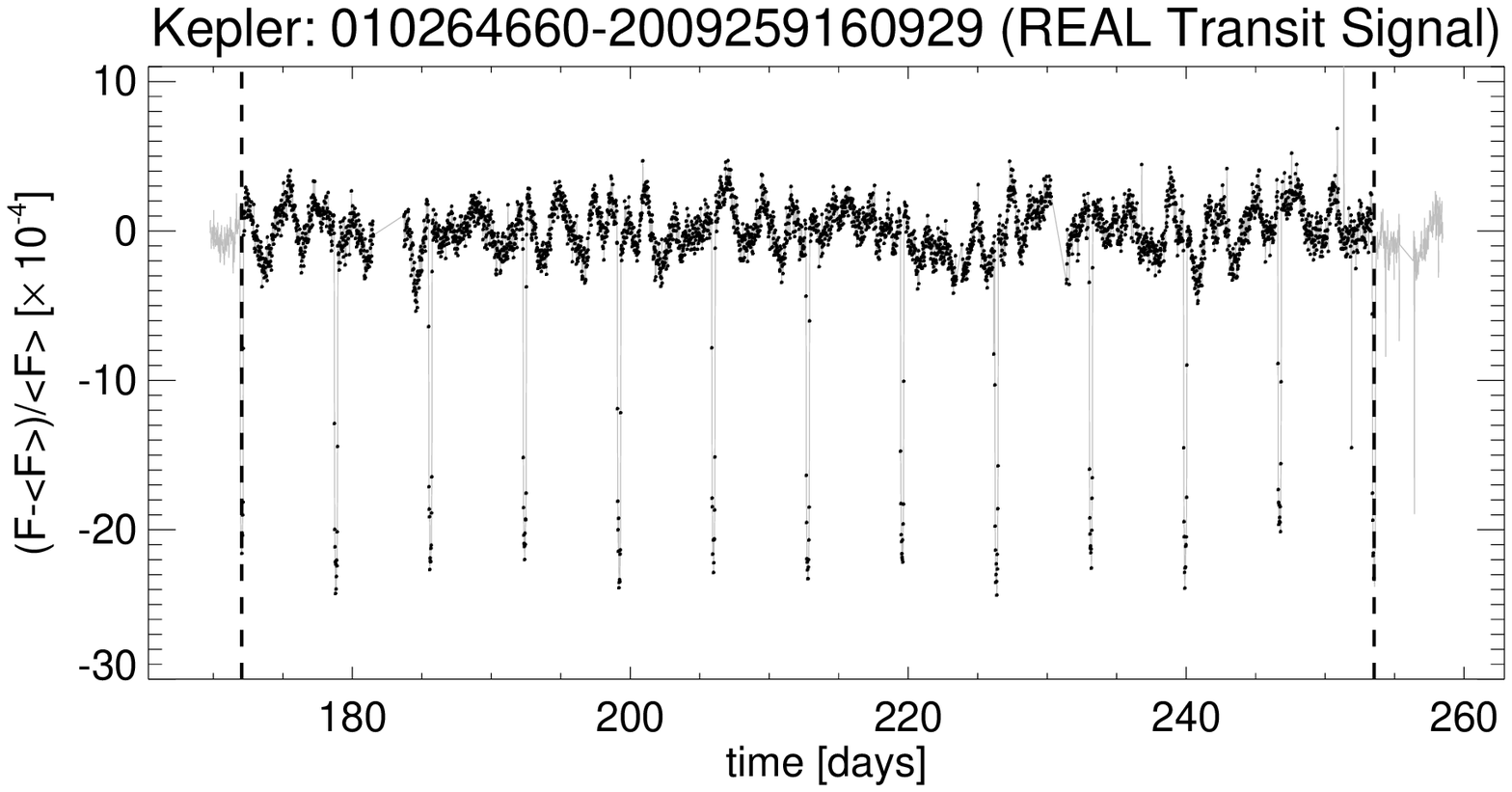}
\includegraphics[width=\columnwidth]{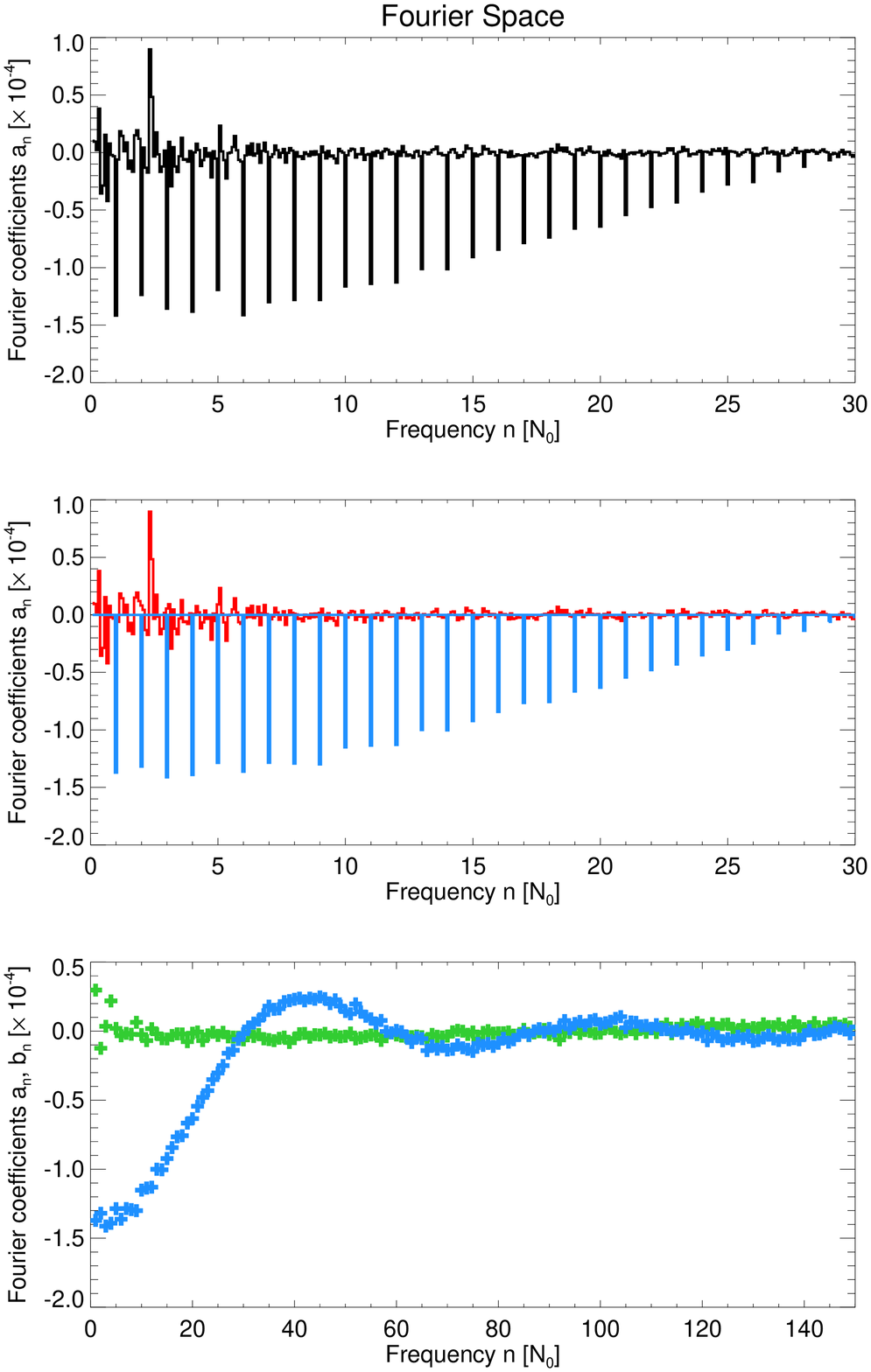}
\caption{Extraction of a planet transit signal from real \emph{Kepler} data using our proposed Fourier method. 
The steps in the method are shown from top to bottom.
\emph{Top}: Rescaled PDCSAP flux from the \emph{Kepler} observed system 010264660-2009259160929.
The \emph{gray line} indicates the full dataset from that quarter, where the \emph{black points} are the measurements we select for our analysis.
As seen, these points are all within the \emph{vertical dashed lines} which are chosen to go through the exact middle of the first and the last planet transit, respectively.
This specific data window greatly improves the separation of signal and noise in Fourier space as described in Section \ref{sec:Choosing_the_Optimal_Data_Window}.
\emph{Top center}: Fourier coefficients $a_{n}$ obtained from representing the selected dataset in a Fourier series, as
shown in equation \eqref{eq:I_Fourier_ex}. The frequency bins with clear narrow peaks pointing toward negative values are the signal bins
and contain a mix of noise and signal. All other bins contain pure noise.
\emph{Bottom center}: The Fourier coefficients $a_{n}$ separated into a noise component $a_{n}^{\mathcal{N}}$ (\emph{red}) and a signal
component $a_{n}^{\mathcal{S}}$ (\emph{blue}). The noise contribution in the signal bins is here estimated by using linear interpolation
as described in Section \ref{sec:Reconstruct_Signal}. This way of removing systematic noise makes our method on average an unbiased estimator of the true transit signal, in
contrast to normal phase folding.
\emph{Bottom}: The full reconstructed signal shown up to the Nyquist frequency.
The \emph{blue} symbols denote $a_{n}^{\mathcal{S}}$ (even part) coefficients, where the \emph{green} symbols denote $b_{n}^{\mathcal{S}}$ (odd part) coefficients.}
\label{fig:method_ill_step_by_step_real_data}
\end{figure}

However, what is usually not appreciated is that the data window actually can be chosen freely
to highlight features of interest in the data. In our case, we can use our knowledge of where the main transits occur in the light curve
to construct a window which will not affect the representation of the signal in Fourier space.
Following this idea, one can show that the minimum overlap between noise and signal generally can be achieved by
choosing a data window such that 
\begin{equation}
\begin{aligned} 
\tau_{i} & = \text{time of \emph{first} transit} \\
\tau_{f} & = \text{time of \emph{last} transit},
\label{eq:time_window}
\end{aligned}
\end{equation}
where the transit time here refers to the time at the center of the main planet transit.
There are two reasons for why this window can significantly improve the reconstruction of the signal:
first, the frequencies related to the signal will now be confined to single frequency bins with
no broadening because $T_{w}$ is now a multiple of the period of the planet. Second, since the time window is now fixed
to start in the middle of a transit, the obtained signal coefficients, $a_{n}^{\mathcal{S}}$ and $b_{n}^{\mathcal{S}}$,
not only describe the odd and even contributions to the data, but also the symmetry of the signal itself.
For example, if we search for a signal which is symmetric around the main transit (as the one shown in Figure \ref{fig:Ill_planet_transit}),
then we know right away that all the $b_{n}^{\mathcal{S}}$ must equal zero. This gives us a perfect separation of noise and signal in
the $b_{n}$ coefficient space, which greatly improves the extraction of the signal.
The top plot in Figure \ref{fig:method_ill_step_by_step_real_data} illustrates the proposed window applied to data from
\emph{Kepler}, where the top center plot shows how this results in Fourier coefficients with distinct and isolated signal peaks.

\subsection{Reconstruction of a Symmetric Transit Signal}\label{sec:Identifying_Noise_and_Signal_in_Fourier_Space}

To make the presentation of the method as clear as possible and to shorten the algebra, we now assume that
the transit shape of the signal is symmetric with respect to the main transit. This is a special case, but in fact represents
a large class of real astrophysical transit phenomena including the main planet transit, secondary eclipse, planetary rings, and trojan-asteroids.
Figure \ref{fig:Ill_planet_transit} shows an example. In section \ref{sec:Reconstruction_of_an_Arbitrary_Transit_Signal} we
describe how to extend the following procedure to reconstruct a signal with arbitrary transit shape.

\subsubsection{Identifying Noise and Signal in Fourier Space}

By assuming the transit shape is symmetric, the signal periodic, and using the window proposed in equation \eqref{eq:time_window},
we then know that only cosine terms will be needed for the reconstruction.
Inserting this case in equation \eqref{eq:I_Fourier_ex} and using the linearity relation between signal and noise shown in equation \eqref{eq:Fourier_coefficients_signal_plus_noise},
we can now express the data described by the cosine terms as
\begin{equation}
I_{c}(t)	 = \sum_{n=1}^{N/2} a_{n}^{\mathcal{N}}{\text{cos} \left( \frac{2{\pi}nt}{T_{w}} \right)} + \sum_{m=1}^{\lfloor N/(2N_{0}) \rfloor} a_{mN_{0}}^{\mathcal{S}}{\text{cos} \left( \frac{2{\pi}(mN_{0})t}{T_{w}} \right)},
\label{eq:I_Fourier_ex_new_win}
\end{equation}
where $N_{0}$ denotes the number of full planet orbits within the data window.
From this equation it is now clear that the noise can contribute at all $n=\{1,2,3... \}$, but the signal only contributes at $n=mN_{0}$ where $m=\{1,2,3...\}$.
In other words, the signal is now constrained to discrete and separate frequency bins in Fourier space.
The fraction these signal bins cover of the total available frequency space
is $\sim 1/(2N_{0})$, which is simply the ratio of the number of signal bins, $N/(2N_0)$, to the total number of bins, $N$.
This means that if $N_{0}=10$ then the signal only appears in $5\%$
of the total available space, i.e., $95\%$ of the modes in Fourier space can be labeled as pure noise.
An example of this signal and noise separation is shown in the middle panels of Figure \ref{fig:method_ill_step_by_step_real_data}.
In the next section we illustrate how the noise contribution can be estimated in the signal bins, which leads to the final extraction of the signal.

\subsubsection{Reconstruct Signal}\label{sec:Reconstruct_Signal}

From using the fixed data window proposed in section \ref{sec:Choosing_the_Optimal_Data_Window},
the signal has now been constrained to only appear in $\sim 1/(2N_{0})$ of the available Fourier space.
However, noise will in general still be present in the bins
where the signal is now localized. The question is if we can remove this noise contribution to improve the reconstruction of the true signal.
We start by using equation \eqref{eq:Fourier_coefficients_signal_plus_noise} to express the signal coefficients as
\begin{equation}
a_{mN_{0}}^{\mathcal{S}} = a_{mN_{0}} - a_{mN_{0}}^{\mathcal{N}},
\end{equation}
where $a_{mN_{0}}$ is obtained directly from data.
If one now can estimate $a_{mN_{0}}^{\mathcal{N}}$, then the underlying signal $a_{mN_{0}}^{\mathcal{S}}$ can be extracted.
The random part of the noise can of course not be removed in individual bins, but we here show that it is possible to correct
for systematic components since these on average appear as smooth features in Fourier space.
The idea is to use the spectral shape of the noise \emph{outside} the signal bins ($n \neq mN_{0}$) to predict
the value \emph{inside} the bins ($n = mN_{0}$).
In other words, we suggest that the systematic contribution to the noise spectrum can on average be estimated by interpolating
across the signal bins in Fourier space, using the regions outside the bins as a baseline. The regions over which we have to interpolate are always
just single frequency bins with unit width, no matter how broad the signal is in real space. The optimal interpolation scheme
depends on the properties of the noise, however, we find that a simple linear interpolation is by far the most robust when using real data.
Using a linear interpolation scheme, the estimate for the true signal coefficients is now given by
\begin{equation}
a_{mN_{0}}^{\mathcal{S}} \approx a_{mN_{0}} - (a_{mN_{0}-1} + a_{mN_{0}+1})/2,
\label{eq:a_mN_S_recon}
\end{equation}
where the last term is just the average of the neighbor bins around $a_{mN_{0}}$. The $a_{x}$ coefficients can directly be found from data by
\begin{equation}
a_{x} = \sum_{j=1}^{N} I(t_{j}) \text{cos} \left(\frac{2{\pi}xt_{j}}{T_{w}}\right),\ 0 \leq t_{j} \leq T_{w},
\label{eq:a_x_coeff}
\end{equation}
where the summation must be performed over the fixed time window as indicated.
If the data is nonuniformly sampled in time, one must use a slightly more general approach for estimating the coefficients. This is further discussed
in the Appendix.

\begin{figure}
\includegraphics[width=\columnwidth]{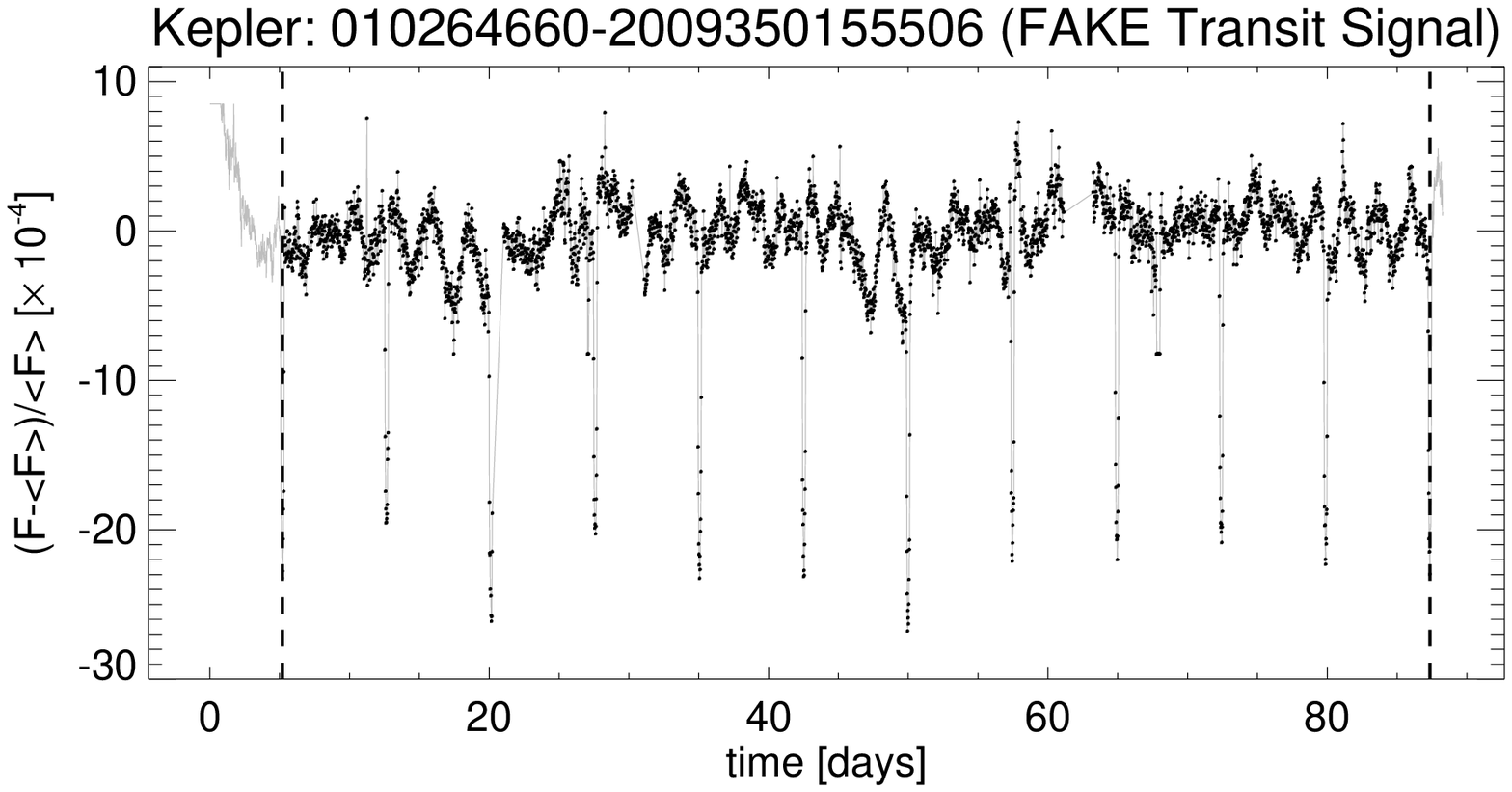}
\includegraphics[width=\columnwidth]{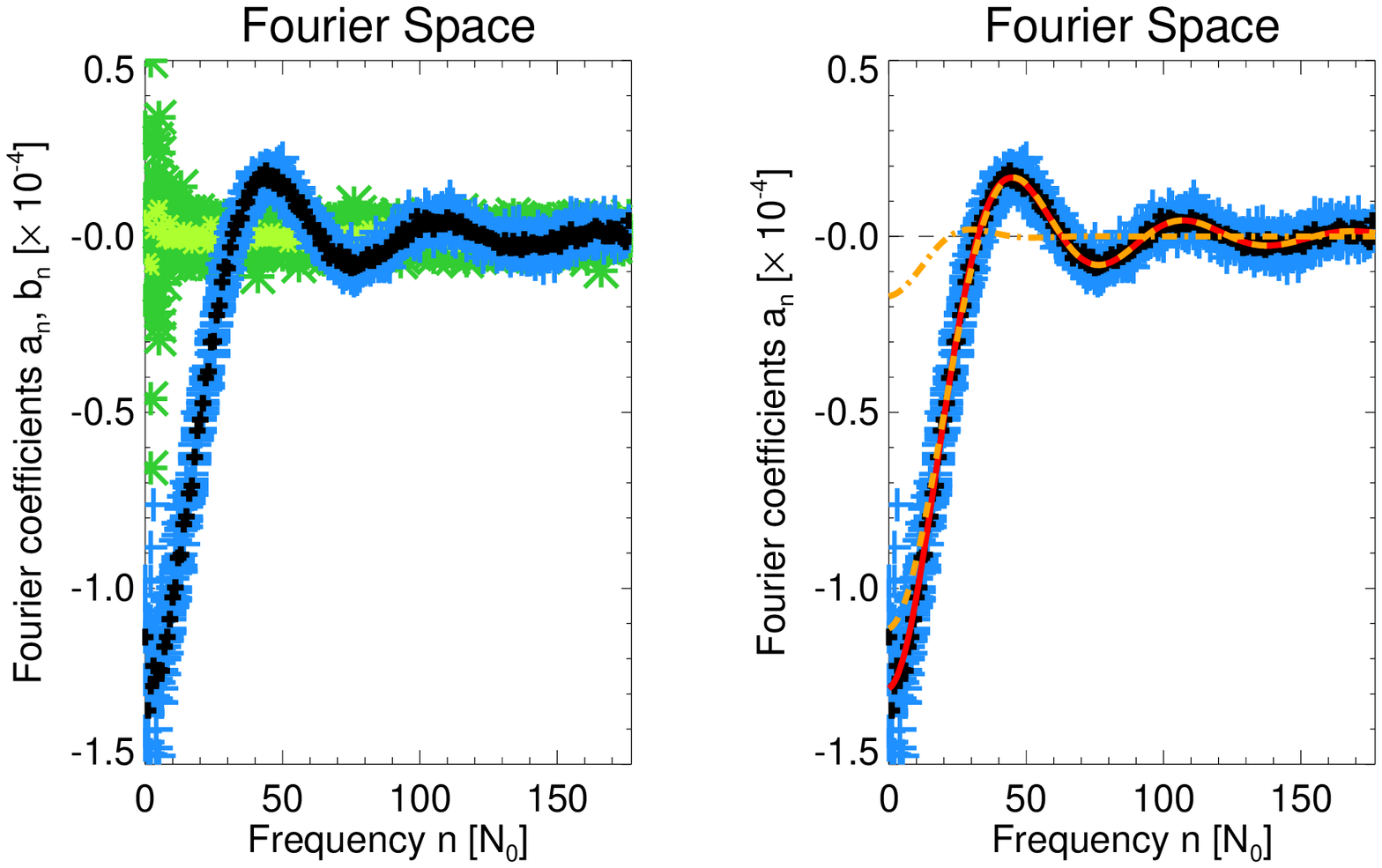}
\includegraphics[width=\columnwidth]{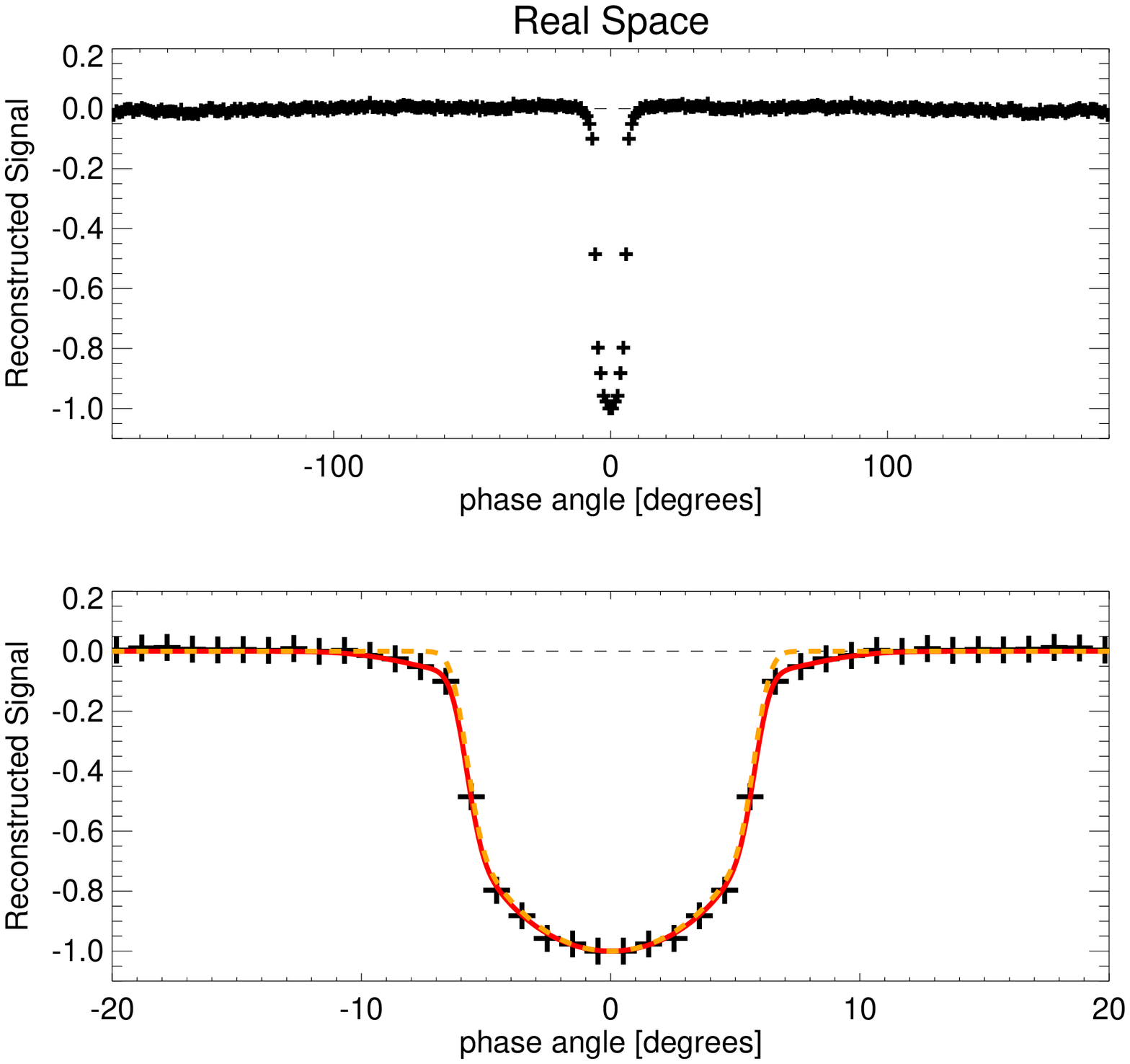}
\caption{Illustration of how well our proposed Fourier method reconstructs a fake signal injected into real
light curves from the \emph{Kepler} telescope. For this example we use $12$ quarters of data from the dataset 010264660.
In addition to the main transit from the planet, we have also added a slightly broader component
with a transit depth of $10\%$ relative to the main transit to emulate a ring system \citep{2004ApJ...616.1193B, 2011ApJ...743...97T}.
The \emph{top plot} shows one of the 12 quarters of data including the fake injected signal.
We use the data inside the matched window illustrated by the \emph{vertical dashed lines}, as described in Section \ref{sec:Choosing_the_Optimal_Data_Window}.
The \emph{top center} panel shows the corresponding Fourier coefficients where $a_{n}$ is shown in \emph{blue} and $b_{n}$ in \emph{green}.
The median values from all 12 quarters are shown in \emph{black} and \emph{light green}, respectively.
In the right plot is shown the Fourier representation of the injected fake signal with a \emph{red solid line},
where the \emph{orange dashed} shows the contribution from the planet and the \emph{orange dashed-dotted} the contribution from the rings.
The \emph{bottom center} plot shows the real space representation of the reconstructed signal using the median estimate.
The \emph{bottom} plot shows a zoom in on the main transit. As seen, our method (black symbols)
reconstructs the true shape (solid red) to a precision where the difference between a signal without rings (dashed line) and with rings (solid line)
clearly can be seen. The same is true for the Fourier representation shown in the top center plot. Any transit shape can be reconstructed by this Fourier
method, as long as the signal is strictly periodic.}
\label{fig:apply_method_fake_data}
\end{figure}

It is often useful to also consider the real space representation of the reconstructed signal.
This representation is simply given by
\begin{equation}
I(t)^{\mathcal{S}} \approx \sum_{m=1}^{\lfloor N/(2N_{0}) \rfloor} a_{mN_{0}}^{\mathcal{S}} \text{cos} \left(\frac{2{\pi}mt}{T_{0}}\right),
\end{equation}
where $T_{0}$ is the orbital time of the planet, and the signal coefficients $a_{mN_{0}}^{\mathcal{S}}$ are given by equation \eqref{eq:a_mN_S_recon}.
The real space representation has the advantage that the different signal components can separate in time along the planetary orbit (see e.g., Figure \ref{fig:apply_method_fake_data}).
However, the drawback is that single outliers among the Fourier coefficients generate global periodic variations, which
makes it very hard to distinguish noise in Fourier space from a possible interesting signal identified in real space. This is
different from the Fourier space representation where all the signal components share the same frequency bins.
A model comparison must therefore be performed
by fitting in Fourier space instead of real space. A blind search for unexpected signals is on the other hand probably best to perform in real
space, where the different parts of the signal often separate and therefore becomes easier to visually identify.

Our estimate of the signal shown in equation \eqref{eq:a_mN_S_recon} is of course not always perfect, but it is expected to
be good on average. This means that if one have several datasets from the same system, 
the stacked set of reconstructed signals in Fourier space will on average be an unbiased estimate of the true signal.
Standard phase folding is not unbiased in the same way, since
the stacking in real space is only optimal for reducing random uncorrelated noise.
A noise contribution with, e.g., a sinusoidal shape has highly correlated values in real space
and is therefore not averaging to zero if divided up in orbital segments and stacked, unless the divisions
are matched in phase and size to the noise shape. However, in Fourier space such a contribution
is basically only occupying a single bin and can therefore in general be
fully removed without touching the signal bins.
Masking out the signal regions in real space and using the data outside to estimate systematic noise
is sometimes possible in phase folding, but highly limited to cases where the location of the signal is already known and
constrained to a small region out of the full orbit. These restrictions do not apply to our method, which makes it possible to
remove noise at all frequencies and reconstruct the full orbital signal.
Our new method is illustrated step by step in Figure \ref{fig:method_ill_step_by_step_real_data} on \emph{Kepler} data.

\subsection{Reconstruction of an Arbitrary Transit Signal}\label{sec:Reconstruction_of_an_Arbitrary_Transit_Signal}
Any function can be described by the sum of an even and an odd function. In this section we have shown how the even part
of the transit shape can be reconstructed. The reconstruction of the odd part is identical to this procedure, 
except that sine terms are then used instead of cosine terms. Any transit shape can therefore be extracted from a dataset with both random and systematic
noise using our proposed method, with the only requirement that the signal must be strictly periodic.


\section{Numerical Example: a Planet with Rings}\label{sec:Numerical_Example}

We here give an example of how well a transit signal can be reconstructed using our Fourier method.
For this illustration we inject a fake signal into $12$ quarters of data from the
\emph{Kepler} object $010264660$. This object already has a planet, as seen in Figure \ref{fig:method_ill_step_by_step_real_data},
which we therefore remove before injecting our fake signal. 
Besides the clean transit signal from the planet itself, we also
include a wider and weaker transit component centered around the planet to emulate a ring system. 
This gives rise to wide wings at ingress and egress, which is one of the main indications for rings \citep{2004ApJ...616.1193B, 2011ApJ...743...97T}.
For illustrative purposes, we chose the injected signal to have a symmetric shape with respect to the main transit. 
The interesting question is now if our method is precise enough to tell if a ring is present or not.

Results are summarized in Figure \ref{fig:apply_method_fake_data}.
There are several ways of how to combine the $12$ datasets, here we use the approach of applying our method to each of the datasets individually.
As seen in the panel showing the Fourier coefficients, each realization has some scatter, but the average of the
$12$ reconstructed sets is consistent with the true signal shape.
As a result, the contribution from rings are easily seen both in Fourier space (top center panel) and in real space (bottom panel).
How to compare the extracted signal to models including proper error propagation, will be discussed in an upcoming paper.


\section{Conclusions}\label{sec:Conclusions}

We present a simple, yet powerful, approach for extracting the shape of periodic transit signals associated with a known planet from an observed light curve.
The method is based on the idea of separating signal and noise in Fourier space, which especially enables us to remove systematic noise components.
We show that this becomes possible if only data within a fixed time window with a width equal to an integer number of orbital periods, is used
in the analysis. In this case, the signal separates out and is described by a significantly smaller number of frequencies than the full dataset.
In addition, the signal frequencies are easily identified in Fourier space, and $\sim 1-(1/N_{0})$ of the total space can
be labeled as pure noise as a result.
For extracting the true transit signal, we suggest that the noise contribution across the frequency bins of the signal can be estimated
by using the regions outside the bins as a baseline. We illustrate this by
successfully extracting a fake transit signal injected into real \emph{Kepler} light curves.

Our proposed method for removing systematic noise does not rely on either input noise models or nearby field stars for calibration.
This makes our method not only model independent, but also fast enough that
several quarters of \emph{Kepler} data can be analyzed in seconds. Furthermore, we can extract the full orbital transit signal without applying
any masking or out-of-transit fitting. Our method has of course some limitations which depend on the data quality and the noise properties,
and one could in general also benefit from combining it with other techniques such as the complementary
trend filtering algorithm (TFA) \citep{2005MNRAS.356..557K}.
This will be explored in an upcoming paper where we apply the method to real data,
to search for rings, trojans and other higher-order photometric effects associated with main orbiting planet.



\acknowledgments{It is a pleasure to thank L. Buchhave, M. Pessah, D. Spergel, G. Bakos, S. Pedersen, C. Huang, 
M. Val-Borro, T. Brandt, J. Hartman for useful conversations,
and especially C. Holcomb for going through an earlier version of the manuscript.
Support for this work was provided by NASA through Einstein Postdoctoral
Fellowship grant number PF4-150127 awarded by the Chandra X-ray Center, which is operated by the
Smithsonian Astrophysical Observatory for NASA under contract NAS8-03060.}


\begin{appendix}

Observationally obtained light curves are often sampled nonuniformly in time and include gaps with no data.
The Fourier coefficients can therefore not be obtained using standard FFT methods, nor by the simple relation shown in equation \eqref{eq:a_x_coeff}.
Instead, one has to perform a joint least square fit of $\text{sine}$ and $\text{cosine}$ functions to the data.
Due to the linearity of the Fourier expansion, this fit is simple and can with great advantage be formulated using matrices. Following this
approach we can now express the data as a matrix multiplication
\begin{equation}
\mathbf{I}_{i} \approx \mathbf{C}_{ni}\mathbf{a}_{n} + \mathbf{S}_{ni}\mathbf{b}_{n} = \mathbf{T}_{ki} \mathbf{R}_{k},
\end{equation}
where $\mathbf{C}_{ni} = \text{cos}(2{\pi}nt_{i}/T_{w})$,  $\mathbf{S}_{ni} = \text{sin}(2{\pi}nt_{i}/T_{w})$, $\mathbf{T}_{ki}$ is their joined matrix,
and $\mathbf{R}_{k} = [\mathbf{a}_{n}, \mathbf{b}_{n}]$ is the vector of coefficients we want to fit for. By minimizing the difference squared
between our expansion and the data, one can now express the solution to the vector $\mathbf{R}_{k}$ as
\begin{equation}
\mathbf{R}_{k} \approx \left( \mathbf{T}^{\text{T}} \mathbf{T} \right)^{-1} \mathbf{T}^{\text{T}} \mathbf{I},
\label{eq:an_bn_matrix_fit}
\end{equation}
where the superscript $\text{T}$ here denotes the transpose. If the data is uniformly sampled then the matrix $\mathbf{T}^{\text{T}} \mathbf{T}$ equals the identity matrix,
and the solution reduces to the simple Fourier relation shown in equation \eqref{eq:a_x_coeff}. Only in this case can a FFT solver be used.
Data from the \emph{Kepler} telescope is very close to being uniformly sampled, however there are many gaps with missing data. One must
therefore use this general fitting approach to keep the frequencies under exact control. All examples in this paper are based on this general fitting.

\end{appendix}


\bibliographystyle{apj}

\begin{thebibliography}{0}
\expandafter\ifx\csname natexlab\endcsname\relax\def\natexlab#1{#1}\fi

\end{thebibliography}


\begin{thebibliography}
\expandafter\ifx\csname natexlab\endcsname\relax\def\natexlab#1{#1}\fi


\bibitem[{{Auvergne} {et~al.}(2009){Auvergne}, {Bodin}, {Boisnard}, {Buey},
  {Chaintreuil}, {Epstein}, {Jouret}, {Lam-Trong}, {Levacher}, {Magnan},
  {Perez}, {Plasson}, {Plesseria}, {Peter}, {Steller}, {Tiph{\`e}ne}, {Baglin},
  {Agogu{\'e}}, {Appourchaux}, {Barbet}, {Beaufort}, {Bellenger}, {Berlin},
  {Bernardi}, {Blouin}, {Boumier}, {Bonneau}, {Briet}, {Butler}, {Cautain},
  {Chiavassa}, {Costes}, {Cuvilho}, {Cunha-Parro}, {de Oliveira Fialho},
  {Decaudin}, {Defise}, {Djalal}, {Docclo}, {Drummond}, {Dupuis}, {Exil},
  {Faur{\'e}}, {Gaboriaud}, {Gamet}, {Gavalda}, {Grolleau}, {Gueguen},
  {Guivarc'h}, {Guterman}, {Hasiba}, {Huntzinger}, {Hustaix}, {Imbert},
  {Jeanville}, {Johlander}, {Jorda}, {Journoud}, {Karioty}, {Kerjean},
  {Lafond}, {Lapeyrere}, {Landiech}, {Larqu{\'e}}, {Laudet}, {Le Merrer},
  {Leporati}, {Leruyet}, {Levieuge}, {Llebaria}, {Martin}, {Mazy}, {Mesnager},
  {Michel}, {Moalic}, {Monjoin}, {Naudet}, {Neukirchner}, {Nguyen-Kim},
  {Ollivier}, {Orcesi}, {Ottacher}, {Oulali}, {Parisot}, {Perruchot},
  {Piacentino}, {Pinheiro da Silva}, {Platzer}, {Pontet}, {Pradines},
  {Quentin}, {Rohbeck}, {Rolland}, {Rollenhagen}, {Romagnan}, {Russ}, {Samadi},
  {Schmidt}, {Schwartz}, {Sebbag}, {Smit}, {Sunter}, {Tello}, {Toulouse},
  {Ulmer}, {Vandermarcq}, {Vergnault}, {Wallner}, {Waultier}, \&
  {Zanatta}}]{2009A&A...506..411A}
{Auvergne}, M., {et~al.} 2009, \aap, 506, 411

\bibitem[{{Bakos} {et~al.}(2002){Bakos}, {L{\'a}z{\'a}r}, {Papp}, {S{\'a}ri},
  \& {Green}}]{2002PASP..114..974B}
{Bakos}, G.~{\'A}., {L{\'a}z{\'a}r}, J., {Papp}, I., {S{\'a}ri}, P., \&
  {Green}, E.~M. 2002, \pasp, 114, 974

\bibitem[{{Barnes} \& {Fortney}(2004)}]{2004ApJ...616.1193B}
{Barnes}, J.~W., \& {Fortney}, J.~J. 2004, \apj, 616, 1193

\bibitem[{{Beaug{\'e}} {et~al.}(2012){Beaug{\'e}}, {Ferraz-Mello}, \&
  {Michtchenko}}]{2012RAA....12.1044B}
{Beaug{\'e}}, C., {Ferraz-Mello}, S., \& {Michtchenko}, T.~A. 2012, Research in
  Astronomy and Astrophysics, 12, 1044

\bibitem[{{Bond} {et~al.}(2004){Bond}, {Udalski}, {Jaroszy{\'n}ski},
  {Rattenbury}, {Paczy{\'n}ski}, {Soszy{\'n}ski}, {Wyrzykowski},
  {Szyma{\'n}ski}, {Kubiak}, {Szewczyk}, {{\.Z}ebru{\'n}}, {Pietrzy{\'n}ski},
  {Abe}, {Bennett}, {Eguchi}, {Furuta}, {Hearnshaw}, {Kamiya}, {Kilmartin},
  {Kurata}, {Masuda}, {Matsubara}, {Muraki}, {Noda}, {Okajima}, {Sako},
  {Sekiguchi}, {Sullivan}, {Sumi}, {Tristram}, {Yanagisawa}, {Yock}, \& {OGLE
  Collaboration}}]{2004ApJ...606L.155B}
{Bond}, I.~A., {et~al.} 2004, \apjl, 606, L155

\bibitem[{{Borucki} {et~al.}(2010){Borucki}, {Koch}, {Basri}, {Batalha},
  {Brown}, {Caldwell}, {Caldwell}, {Christensen-Dalsgaard}, {Cochran},
  {DeVore}, {Dunham}, {Dupree}, {Gautier}, {Geary}, {Gilliland}, {Gould},
  {Howell}, {Jenkins}, {Kondo}, {Latham}, {Marcy}, {Meibom}, {Kjeldsen},
  {Lissauer}, {Monet}, {Morrison}, {Sasselov}, {Tarter}, {Boss}, {Brownlee},
  {Owen}, {Buzasi}, {Charbonneau}, {Doyle}, {Fortney}, {Ford}, {Holman},
  {Seager}, {Steffen}, {Welsh}, {Rowe}, {Anderson}, {Buchhave}, {Ciardi},
  {Walkowicz}, {Sherry}, {Horch}, {Isaacson}, {Everett}, {Fischer}, {Torres},
  {Johnson}, {Endl}, {MacQueen}, {Bryson}, {Dotson}, {Haas}, {Kolodziejczak},
  {Van Cleve}, {Chandrasekaran}, {Twicken}, {Quintana}, {Clarke}, {Allen},
  {Li}, {Wu}, {Tenenbaum}, {Verner}, {Bruhweiler}, {Barnes}, \&
  {Prsa}}]{2010Sci...327..977B}
{Borucki}, W.~J., {et~al.} 2010, Science, 327, 977

\bibitem[{{Charbonneau} {et~al.}(2000){Charbonneau}, {Brown}, {Latham}, \&
  {Mayor}}]{2000ApJ...529L..45C}
{Charbonneau}, D., {Brown}, T.~M., {Latham}, D.~W., \& {Mayor}, M. 2000, \apjl,
  529, L45

\bibitem[{{Dyudina} {et~al.}(2005){Dyudina}, {Sackett}, {Bayliss}, {Seager},
  {Porco}, {Throop}, \& {Dones}}]{2005ApJ...618..973D}
{Dyudina}, U.~A., {Sackett}, P.~D., {Bayliss}, D.~D.~R., {Seager}, S., {Porco},
  C.~C., {Throop}, H.~B., \& {Dones}, L. 2005, \apj, 618, 973

\bibitem[{{Heller}(2014)}]{2014ApJ...787...14H}
{Heller}, R. 2014, \apj, 787, 14

\bibitem[{{Henry} {et~al.}(2000){Henry}, {Marcy}, {Butler}, \&
  {Vogt}}]{2000ApJ...529L..41H}
{Henry}, G.~W., {Marcy}, G.~W., {Butler}, R.~P., \& {Vogt}, S.~S. 2000, \apjl,
  529, L41

\bibitem[{{Hippke}(2015)}]{2015arXiv150205033H}
{Hippke}, M. 2015, ArXiv e-prints

\bibitem[{{Hui} \& {Seager}(2002)}]{2002ApJ...572..540H}
{Hui}, L., \& {Seager}, S. 2002, \apj, 572, 540

\bibitem[{{Kenworthy} \& {Mamajek}(2015)}]{2015ApJ...800..126K}
{Kenworthy}, M.~A., \& {Mamajek}, E.~E. 2015, \apj, 800, 126

\bibitem[{{Kipping} {et~al.}(2012){Kipping}, {Bakos}, {Buchhave},
  {Nesvorn{\'y}}, \& {Schmitt}}]{2012ApJ...750..115K}
{Kipping}, D.~M., {Bakos}, G.~{\'A}., {Buchhave}, L., {Nesvorn{\'y}}, D., \&
  {Schmitt}, A. 2012, \apj, 750, 115

\bibitem[{{Kov{\'a}cs} {et~al.}(2005){Kov{\'a}cs}, {Bakos}, \&
  {Noyes}}]{2005MNRAS.356..557K}
{Kov{\'a}cs}, G., {Bakos}, G., \& {Noyes}, R.~W. 2005, \mnras, 356, 557

\bibitem[{{Laughlin} \& {Chambers}(2002)}]{2002AJ....124..592L}
{Laughlin}, G., \& {Chambers}, J.~E. 2002, \aj, 124, 592

\bibitem[{{Laughlin} \& {Lissauer}(2015)}]{2015arXiv150105685L}
{Laughlin}, G., \& {Lissauer}, J.~J. 2015, ArXiv e-prints

\bibitem[{{Lissauer} {et~al.}(2014){Lissauer}, {Dawson}, \&
  {Tremaine}}]{2014Natur.513..336L}
{Lissauer}, J.~J., {Dawson}, R.~I., \& {Tremaine}, S. 2014, \nat, 513, 336

\bibitem[{{Mayor} {et~al.}(2014){Mayor}, {Lovis}, \&
  {Santos}}]{2014Natur.513..328M}
{Mayor}, M., {Lovis}, C., \& {Santos}, N.~C. 2014, \nat, 513, 328

\bibitem[{{Mayor} \& {Queloz}(1995)}]{1995Natur.378..355M}
{Mayor}, M., \& {Queloz}, D. 1995, \nat, 378, 355

\bibitem[{{Moldovan} {et~al.}(2010){Moldovan}, {Matthews}, {Gladman}, {Bottke},
  \& {Vokrouhlick{\'y}}}]{2010ApJ...716..315M}
{Moldovan}, R., {Matthews}, J.~M., {Gladman}, B., {Bottke}, W.~F., \&
  {Vokrouhlick{\'y}}, D. 2010, \apj, 716, 315

\bibitem[{{Perryman}(2011)}]{2011exha.book.....P}
{Perryman}, M. 2011, {The Exoplanet Handbook}

\bibitem[{{Pollacco} {et~al.}(2006){Pollacco}, {Skillen}, {Collier Cameron},
  {Christian}, {Hellier}, {Irwin}, {Lister}, {Street}, {West}, {Anderson},
  {Clarkson}, {Deeg}, {Enoch}, {Evans}, {Fitzsimmons}, {Haswell}, {Hodgkin},
  {Horne}, {Kane}, {Keenan}, {Maxted}, {Norton}, {Osborne}, {Parley}, {Ryans},
  {Smalley}, {Wheatley}, \& {Wilson}}]{2006PASP..118.1407P}
{Pollacco}, D.~L., {et~al.} 2006, \pasp, 118, 1407

\bibitem[{{Sanchis-Ojeda} {et~al.}(2014){Sanchis-Ojeda}, {Rappaport}, {Winn},
  {Kotson}, {Levine}, \& {El Mellah}}]{2014ApJ...787...47S}
{Sanchis-Ojeda}, R., {Rappaport}, S., {Winn}, J.~N., {Kotson}, M.~C., {Levine},
  A., \& {El Mellah}, I. 2014, \apj, 787, 47

\bibitem[{{Sidis} \& {Sari}(2010)}]{2010ApJ...720..904S}
{Sidis}, O., \& {Sari}, R. 2010, \apj, 720, 904

\bibitem[{{Simon} {et~al.}(2012){Simon}, {Szab{\'o}}, {Kiss}, \&
  {Szatm{\'a}ry}}]{2012MNRAS.419..164S}
{Simon}, A.~E., {Szab{\'o}}, G.~M., {Kiss}, L.~L., \& {Szatm{\'a}ry}, K. 2012,
  \mnras, 419, 164

\bibitem[{{Swain} {et~al.}(2010){Swain}, {Deroo}, {Griffith}, {Tinetti},
  {Thatte}, {Vasisht}, {Chen}, {Bouwman}, {Crossfield}, {Angerhausen},
  {Afonso}, \& {Henning}}]{2010Natur.463..637S}
{Swain}, M.~R., {et~al.} 2010, \nat, 463, 637

\bibitem[{{Turner} {et~al.}(2014){Turner}, {Fromang}, {Gammie}, {Klahr},
  {Lesur}, {Wardle}, \& {Bai}}]{2014prpl.conf..411T}
{Turner}, N.~J., {Fromang}, S., {Gammie}, C., {Klahr}, H., {Lesur}, G.,
  {Wardle}, M., \& {Bai}, X.-N. 2014, Protostars and Planets VI, 411

\bibitem[{{Tusnski} \& {Valio}(2011)}]{2011ApJ...743...97T}
{Tusnski}, L.~R.~M., \& {Valio}, A. 2011, \apj, 743, 97

\bibitem[{{Udalski} {et~al.}(2002){Udalski}, {Paczynski}, {Zebrun},
  {Szymanski}, {Kubiak}, {Soszynski}, {Szewczyk}, {Wyrzykowski}, \&
  {Pietrzynski}}]{2002AcA....52....1U}
{Udalski}, A., {et~al.} 2002, \actaa, 52, 1

\bibitem[{{Udry} \& {Santos}(2007)}]{2007ARA&A..45..397U}
{Udry}, S., \& {Santos}, N.~C. 2007, \araa, 45, 397

\bibitem[{{Wolszczan} \& {Frail}(1992)}]{1992Natur.355..145W}
{Wolszczan}, A., \& {Frail}, D.~A. 1992, \nat, 355, 145

\bibitem[{{Zellem} {et~al.}(2014){Zellem}, {Griffith}, {Deroo}, {Swain}, \&
  {Waldmann}}]{2014ApJ...796...48Z}
{Zellem}, R.~T., {Griffith}, C.~A., {Deroo}, P., {Swain}, M.~R., \& {Waldmann},
  I.~P. 2014, \apj, 796, 48

\bibitem[{{Zhu} {et~al.}(2014){Zhu}, {Huang}, {Zhou}, \&
  {Lin}}]{2014ApJ...796...67Z}
{Zhu}, W., {Huang}, C.~X., {Zhou}, G., \& {Lin}, D.~N.~C. 2014, \apj, 796, 67


\end{thebibliography}


\end{document}